\documentclass[preprintnumbers,byrevtex]{revtex4}
\usepackage{amsmath,amsfonts,amssymb,amscd,amsxtra,amsthm}
\usepackage{graphicx}
\usepackage{bm}
\usepackage{epstopdf}
\usepackage{multirow}
\begin{document}
\preprint{KIAS-P12071}
\title{Electrical conductivity of quark matter at finite $T$}
\author{Seung-il Nam}
\email[E-mail: ]{sinam@kias.re.kr}
\affiliation{School of Physics, Korea Institute for Advanced Study (KIAS), Seoul 130-722, Republic of Korea}
\date{\today}
\begin{abstract}
In this talk, I present the recent theoretical results on the electrical  conductivity (EC) $\sigma$ of quark matter, using the Kubo formula at finite temperature and zero quark density $(T\ne0,\mu=0)$ in the presence of an external strong magnetic field. The dilute instanton-liquid model with the caloron distribution is taken into account. It turns out that $\sigma\approx(0.02\sim0.15)\,\mathrm{fm}^{-1}$ for $T=(0\sim400)$ MeV with the relaxation time $\tau=(0.3\sim0.9)$ fm. EC is parameterized as $\sigma/T\approx (0.46,0.77,1.08,1.39)\,C_\mathrm{EM}$ for $\tau=(0.3,0.5,0.7,0.9)$ fm, respectively. These results are well compatible with other theoretical estimations and show almost negligible effects from the magnetic field. The soft photon emission rate from the quark-gluon plasma is discussed as well.  
\end{abstract}
\maketitle
\section{Introduction}
The observation of the strong magnetic field in the peripheral heavy-ion collision at Relativistic Heavy Ion Collider (RHIC) of BNL~\cite{Voloshin:2008jx} triggered abundant related research works~\cite{Kharzeev:2004ey,Buividovich:2009wi,Fukushima:2008xe,Kharzeev:2010gd,Nam:2011vn,Ding:2010ga,Kerbikov:2012vp}. The transport coefficients for the hot and/or dense matter play an important role in general as well, since they determine the physical properties of the matter, being studied by the Kubo formula~\cite{BOOKEM}. In the present work, I would like to investigate the electrical  conductivity (EC) $\sigma$, relating to the vector-current correlation (VCC) in the presence of the external static magnetic field $\bm{B}$.  EC was investigated in the hot phase of the QCD plasma and extracted from a quenched SU($N_c$) lattice QCD (LQCD) in Refs.~\cite{Ding:2010ga,Gupta:2003zh}. Beside LQCD, The authors explored EC using  the Green-function method~\cite{Kerbikov:2012vp}. Note that EC is deeply related to the thermal dilepton production from the quark-gluon plasma (QGP)~\cite{Harada:2006hu}. To this end, I will use the dilute instanton-liquid model~\cite{Shuryak:1981ff,Diakonov:1983hh}, modified by the caloron~\cite{Diakonov:1988my,Nam:2009nn}, resulting in that the instanton size $\bar{\rho}$ becomes a smoothly decreasing function of $T$, signaling weakening nonperturbative effects of QCD. From the numerical results, it turns out that $\sigma\approx(0.02\sim0.15)\,\mathrm{fm}^{-1}$ for $T=(0\sim400)$ MeV with the relaxation time $\tau=(0.3\sim0.9)$ fm. In addition, The parameterization of EC is given as $\sigma/T\approx (0.46,0.77,1.08,1.39)\,C_\mathrm{EM}$ for $\tau=(0.3,0.5,0.7,0.9)$ fm, respectively. These results are well compatible with other theoretical estimations and show almost negligible effects from the magnetic field. The soft photon emission rate from the quark-gluon plasma is discussed as well.
\section{Theoretical framework}
First, EC can be defined in Euclidean space from the Kubo formula~\cite{Kerbikov:2012vp}:
\begin{equation}
\label{eq:EC1}
\sigma_{\mu\nu}(p)=-\sum_f\frac{e^2_f}{w_p}
\int\frac{d^4k}{(2\pi)^4}\mathrm{Tr}_{c,\gamma}
\left[S(k)\gamma_\mu S(k+p)\gamma_\nu\right]_A.
\end{equation}
Here, $e_f$ stands for the electrical  charge for a light-flavor ($f$) quark. $w_p$ indicates the Matsubara frequency for the momentum $p$ for $\sigma$, being proportional to $2\pi T$. $\mathrm{Tr}_{c,\gamma}$ is assigned as the trace over the color and Lorentz indices. In order to evaluate Eq.~(\ref{eq:EC1}), the dilute instanton-liquid mode is employed~\cite{Shuryak:1981ff,Diakonov:1983hh}. In Euclidean space, I define the effective chiral action (EChA) of the model:
\begin{equation}
\label{eq:ECA}
\mathcal{S}_\mathrm{eff}=-\mathrm{Sp}_{c,f,\gamma}\ln
\left[i\rlap{\,\,/}{D}-iM(D^2) \right],
\end{equation}
where $\mathrm{Sp}_{c,f,\gamma}$ stands for the functional trace, while $M(D^2)$ for the momentum-dependent effective quark mass with the U(1) covariant derivative $D_\mu$.The external electromagnetic (EM) field is induced to EChA via the Schwinger method~\cite{Schwinger:1951nm}. From EChA, one can derive the light-quark propagator under the external EM field in the momentum space as~\cite{Nam:2009jb}
\begin{equation}
\label{eq:PRO}
S(K)\approx
\frac{\rlap{\,/}{K}+i[M_k+\frac{1}{2}\tilde{M}_k(\sigma\cdot F)]}{k^2+M^2_k},
\,\,\,\,
M_k=M_0\left[\frac{2}{2+\bar{\rho}^2k^2} \right]^2,\,\,\,\,
\tilde{M}_k=-\frac{8M_0\bar{\rho}^2}{(2+\bar{\rho}^2k^2)^3},
\end{equation}
where $K_\mu\equiv k_\mu+e_fA_\mu$ and $\sigma\cdot F\equiv \sigma_{\mu\nu}F^{\mu\nu}$. $M_0$ is determined as about $350$ MeV for $1/\bar{\rho}\approx600$ MeV. Taking into account the $T$ dependence of $M$ as in Ref.~\cite{Nam:2009nn}, I can write the ($|k|,T$)-dependent $M$ as follows:
\begin{equation}
\label{eq:momo}
M_k=M_0\left[\frac{\sqrt{n(T)}\,\bar{\rho}^2(T)}
{\sqrt{n(0)}\,\bar{\rho}^2(0)}\right]
\left[\frac{2}{2+k^2\,\bar{\rho}^2(T)} \right]^2.
\end{equation}
Considering all the ingredients discussed so far and performing the fermionic Matsubara formula, I arrive at
\begin{eqnarray}
\label{eq:EC4}
\sigma_\perp&=&\sum_fe^2_fN_c\tau
\Bigg\{\int\frac{d^3\bm{k}}{(2\pi)^3}
F^2_{\bm{k}}(\bm{k}^2)\left[
\frac{\mathrm{tanh}\left(\pi \tau E_{\bm{k}} \right)}{E_{\bm{k}}}\right] 
\cr
&+&\frac{\tau\pi}{2}\int\frac{d^3\bm{k}}{(2\pi)^3}
F^2_{\bm{k}}(\bm{k}^2)\left[\frac{\mathrm{sech}^2\left(\pi\tau E_{\bm{k}}\right)}{E^3_{\bm{k}}}
\left[\frac{\mathrm{sinh}\left(2\pi\tau E_{\bm{k}} \right)}{2\pi\tau}-E_{\bm{k}}\right]
\right]\left[M^2_{\bm{k}}-4\tilde{M}^2_{\bm{k}}(e_fB_0)^2 \right]\Bigg\}.
\end{eqnarray}
The energy of the quark is given by $E_{\bm{k}}=(\bm{k}^2+M^2_{\bm{k}})^{1/2}$. I note that I inserted $F^2_{\bm{k}}(\bm{k}^2)$ in the integrals in Eq.~(\ref{eq:EC4}) to tame the UV divergence smoothly in integrating over $\bm{k}$, instead of setting a three-dimensional cutoff. $\sigma_\parallel$ can be easily obtained by putting $B_0=0$ in Eq.~(\ref{eq:EC4}). Details of the present theoretical framework are given in Refs.~\cite{Nam:2009nn,Nam:2009jb,Nam:2012sg}.

\section{Numerical results}
\begin{figure}[t]
\begin{tabular}{cc}
\includegraphics[width=8cm]{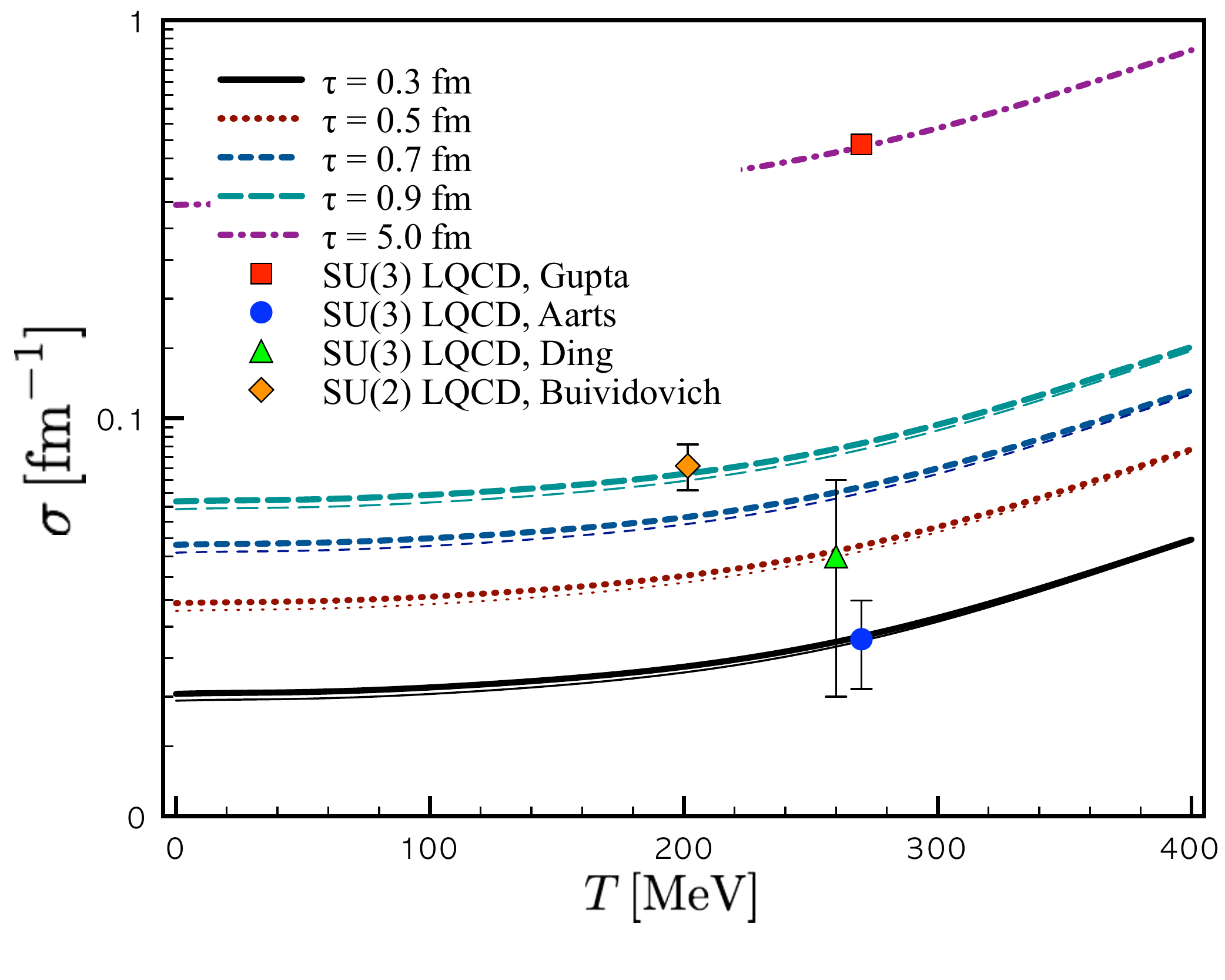}
\includegraphics[width=8cm]{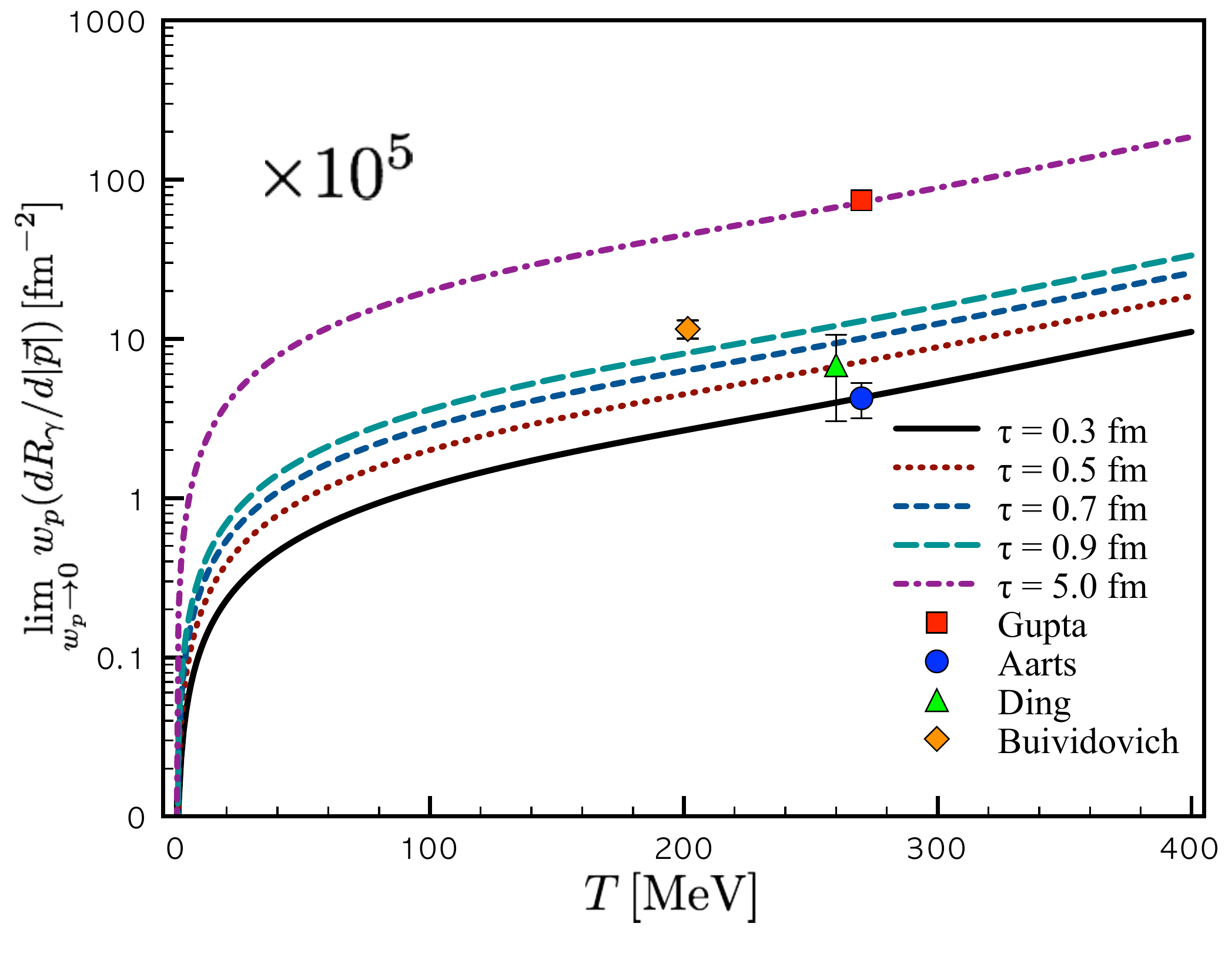}
\end{tabular}
\caption{(Color online) Left: EC $\sigma$ as functions of temperature $T$ for different relaxation times $\tau=(0.3\sim0.9)$ fm. The thick line indicate the case with $B_0=0$ ($\sigma_\parallel$), whereas the thin ones for $B_0=m^2_\pi\times10$ ($\sigma_\perp$). SU($N_c$) LQCD estimations are taken from Refs.~\cite{Gupta:2003zh} (Gupta), ~\cite{Aarts:2007wj} (Aarts),~ \cite{Ding:2010ga} (Ding), and ~\cite{Buividovich:2010tn} (Buividovich). Right: Soft photon emission rate $\mathcal{R}_\gamma$ in Eq.~(XXX) as functions of $T$ for different $\tau$'s in the same manner with the left panel.  For all the curves in the left and right panels, I have chosen $T_c\approx180$ MeV as a trial.}        
\label{FIG34}
\end{figure}

The numerical results for EC and the comparisons with other theoretical results are presented. In the left panel of figure~\ref{FIG34}, I show the numerical results of $\sigma_\perp$ (thick) and $\sigma_\parallel$ (thin) for different $\tau$ values, $\tau=(0.1,0.3,0.5,0.9)$ fm in (solid, dot, dash, dot-dash) lines, respectively, as functions of $T$. The external magnetic field is chosen to be $B_0=m^2_\pi\times10$, where $\bm{B}=(0,0,B_0)$, as a trial. Note that this value of $B_0$ is much stronger than that observed in the RHIC experiment~\cite{Voloshin:2008jx}. EC shows a rapidly increasing curve with respect to $T$ and show the obvious increases beyond $T\approx200$ MeV. By comparing those cases with and without $B_0$, one sees that the effect from the external magnetic field is negligible and only relatively effective in the low-$T$ region $T\lesssim200$ MeV, i.e. $\sigma_\perp\approx\sigma_\parallel\equiv\sigma$. Note that $\sigma$ values for some typical temperatures are given in table~\ref{TABLE1}, in which one can easily see that $\sigma$ is rather linear for $T\lesssim150$ MeV, and increases monotonically after it. At $T_c\approx180$ MeV, which is close to the transition temperature of QCD, one obtains $\sigma=(0.023,0.039,0.054,0.070)\,\mathrm{fm}^{-1}$ for $\tau=(0.1,0.3,0.5,0.9)$ fm. The recent LQCD simulations~\cite{Aoki:2006br,Aoki:2009sc,Borsanyi:2010bp,Bazavov:2011nk}, give the transition temperature as $T_c\approx155$ MeV, which is lower than those in Refs.~\cite{Maezawa:2007fd,Ali Khan:2000iz}. Taking  $T_c\approx155$ MeV, $\sigma$ becomes $(0.022,0.037,0.052,0.067)\,\mathrm{fm}^{-1}$ for $\tau=(0.1,0.3,0.5,0.9)$ fm. Hence, I conclude that only small changes are observed for $\sigma$ for $T_c\lesssim200$ MeV as shown in the left panel of figure~\ref{FIG34}.

\begin{table}[h]
\begin{center}
\begin{tabular}{c|c|c|c|c|c}
&\hspace{0.5cm}$T=0$\hspace{0.5cm}
&$T=100$ MeV&$T=200$ MeV&$T=300$ MeV&$T=400$ MeV\\
\hline
$\tau=0.3$ fm&$0.020$&$0.021$&$0.024$&$0.031$&$0.049$\\
$\tau=0.5$ fm&$0.034$&$0.036$&$0.040$&$0.053$&$0.083$\\
$\tau=0.7$ fm&$0.048$&$0.050$&$0.056$&$0.074$&$0.116$\\
$\tau=0.9$ fm&$0.062$&$0.064$&$0.072$&$0.095$&$0.149$\\
\end{tabular}
\caption{Typical values of $\sigma$ [fm$^{-1}$] at $B_0=0$ 
for different $T$ and $\tau$ values.}
\label{TABLE1}
\end{center}
\end{table}

For practical applications as in the LQCD simulations~\cite{Gupta:2003zh,Aarts:2007wj}, it is quite convenient to parameterize EC as follows:
\begin{equation}
\label{eq:PARAEC}
\sigma(T)=C_\mathrm{EM}\sum_{m=1}\mathcal{C}_m T^m,\,\,\,\,\frac{\mathcal{C}_m}{\mathrm{fm^{m-1}}}\in\mathcal{R},
\end{equation}
where $C_\mathrm{EM}$ is defined as $\sum_fe^2_f\approx0.051$ for the SU(2) light-flavor sector. The coefficients computed up to $m=3$ are given in table~\ref{TABLE2}. As understood from the coefficients, EC becomes almost linearly as functions of $T$, i.e. $|\mathcal{C}_{2,3}|\sim0$. Hence, one can approximate them as $\sigma/T\approx (0.46,0.77,1.08,1.39)\,C_\mathrm{EM}$ for $\tau=(0.3,0.5,0.7,0.9)$ fm, respectively, to a certain extent.
\begin{table}[b]
\begin{center}
\begin{tabular}{c||c|c|c|c}
&$\tau=0.3$ fm&$\tau=0.5$ fm&$\tau=0.7$ fm&$\tau=0.9$ fm\\
\hline
$\mathcal{C}_1$&$0.46$&$0.77$&$1.08$&$1.39$\\
$\mathcal{C}_2$ [fm] &$4.00\times10^{-6}$&$6.66
\times10^{-6}$&$9.33\times10^{-6}$&$1.20\times10^{-6}$\\
$\mathcal{C}_3$ [fm$^2$]&$-4.87\times10^{-5}$&$-4.87
\times10^{-6}$&$-4.88\times10^{-5}$&$-4.88\times10^{-5}$\\
\end{tabular}
\caption{The coefficients $\mathcal{C}_{1,2,3}$ for different $\tau$ values.}
\label{TABLE2}
\end{center}
\end{table}

In Refs.~\cite{Gupta:2003zh} and~\cite{Aarts:2007wj}, employing the SU(3) quenched LQCD simulations, it was estimated that $\sigma/T=7C_\mathrm{EM}$ for $1.5<T/T_c<3$ and $\sigma/T=(0.4\pm0.1)C_\mathrm{EM}$ for $T/T_c\approx1.5$, respectively. Note that there is one order difference between these $\sigma$ values, although the temperature ranges are not overlapped. In the left panel of figure~\ref{FIG34}, I depict these two LQCD values from Ref.~\cite{Gupta:2003zh} (square) and Ref.~\cite{Aarts:2007wj} (circle), using $T_c\approx180$ MeV as a trial, although the transition temperatures are slightly higher than this value in general in the quenched LQCD simulations. It is shown that the data point from Ref.~\cite{Aarts:2007wj} is well consistent with our results for $\tau\approx0.3$ fm. In contrast, the data point from Ref.~\cite{Gupta:2003zh} for $T=270$ MeV is much larger than ours for $\tau=(0.3\sim0.9)$ fm. I verified that, in order to reproduce it, $\tau$ becomes about $5$ fm in our model calculation as shown in the left panel of figure~\ref{FIG34} in the dot-dash line. In Ref.~\cite{Tuchin:2010vs}, the characteristic $\tau$ was estimated using Ref.~\cite{Gupta:2003zh}, resulting in  $\sim(2.2\,T/T_c)$ fm with a conservative estimate of the QGP medium size. Taking $T\approx270\,\mathrm{MeV}=T_c\times1.5$ MeV, it is given that $\tau\approx5$ fm, which is in good agreement with our model results as depicted in the left panel of figure~\ref{FIG34}. Comparably, at $T\approx1.45\,T_c$, it was suggested that $\sigma/T\approx (1/3\sim1)C_\mathrm{EM}$ in Ref.~\cite{Ding:2010ga}. If I choose $T_c\approx180$ MeV again, this result provides that $\sigma=(0.022\sim0.067)\,\mathrm{fm}^{-1}$, which is drawn in the left panel of figure~\ref{FIG34} (triangle) and it corresponds to $\tau\approx(0.3\sim0.7)$ fm in comparison with our results. The typical time scale of $\tau$ was given by $\tau T=0.5$, giving $\tau\approx0.38$ fm at $T\approx1.45\,T_c$~\cite{Ding:2010ga}. Interestingly, this time scale is compatible with ours. 

In Ref.~\cite{Buividovich:2010tn}, the quenched SU(2) LQCD was performed, and EC was also estimated as $\sigma=(0.076\pm0.010)\,\mathrm{fm}^{-1}$ at $T=350$ MeV with the transition temperature $\sim313$ MeV, due to $T=1.12\,T_c$. To be consistent with others using $T_c=180$ MeV as above, I depict the data point of Ref.~\cite{Buividovich:2010tn} at $T=1.12\times180\,\mathrm{MeV}\approx202$ MeV in the left panel of figure~\ref{FIG34} (diamond), although it was evaluated at $T=350$ MeV. Being different from other LQCD data, Ref.~\cite{Buividovich:2010tn} presented the longitudinal and transverse components of $\sigma$ separately in the presence of the external magnetic field. Those LQCD data showed that $\sigma_\perp\approx\sigma_\parallel$ beyond $T_c$ for arbitrary values of $B_0$, whereas $\sigma_\perp\ne\sigma_\parallel$ at $T=0$ and the difference between them is enhanced by increasing $B_0$. Qualitatively, this observation of the LQCD results are consistent with ours as indicated by the thick and thin lines in the left panel of figure~\ref{FIG34}. In our calculations, $\sigma_\parallel$ (thin) is smaller than $\sigma_\perp$, mainly due to that the negative sign in front of the term $\propto(e_fB_0)^2$ in Eq.~(\ref{eq:EC4}) in the vicinity of $T\approx0$. On the contrary, $\sigma_\parallel$ gets larger than $\sigma_\perp$ at $T=0$ in Ref.~\cite{Buividovich:2010tn}. 

Beside the LQCD data, one has several theoretical estimations for EC via effective approaches using the Green-function method~\cite{Kerbikov:2012vp} and ChPT~\cite{FernandezFraile:2009mi}. In Ref.~\cite{Kerbikov:2012vp}, EC was computed for finite temperature and quark density, $T=100$ MeV and $\mu=400$ MeV, which corresponds to the future heavy-ion collision facilities (FAIR, NICA). By choosing $\tau=0.9$ fm, it was given that $\sigma\approx0.04\,\mathrm{fm}^{-1}$. Note that this value corresponds to $\tau=(0.5\sim0.7)$ in our results for zero density. In other words, by increasing the quark density, EC decreases at a certain temperature, as expected.

Finally, I would like to estimate the (differential) soft photon ($w_p\to0$) emission rate from QGP for the dilepton decay rates which is related to EC as follows~\cite{Gupta:2003zh,Ding:2010ga}: 
\begin{equation}
\label{eq:SPER}
\mathcal{R}_\gamma\equiv
\lim_{w_p\to0}w_p\frac{dR_\gamma}{d^3|\bm{p}|}=\frac{3\alpha_\mathrm{EM}}{2\pi^2}\sigma T.
\end{equation}
The numerical results for $\mathcal{R}_\gamma$ in Eq.~(\ref{eq:SPER}) are given in the right panel of figure~\ref{FIG34} for different $\tau$, being similar to the left panel of figure~\ref{FIG34}. The LQCD results are also depicted there. It turns out that the value of $\mathcal{R}_\gamma$ increases rapidly near $T=0$. Beyond that, the slope of $\mathcal{R}_\gamma$ becomes rather flat for $T$. 
\section{Summary and conclusion}
EC is an increasing function of $T$ and depends on $\tau$ of the quark matter. The effective quark mass was modified into a decreasing function of $T$ and $|\bm{k}|$. Typically, it turns out that $\sigma\approx(0.02\sim0.15)\,\mathrm{fm}^{-1}$ for $T=(0\sim400)$ MeV with the relaxation time $\tau=(0.3\sim0.9)$ fm. Recent LQCD data are well reproduced for $\tau=(0.3\sim0.7)\,\mathrm{fm}$ for a wide $T$ range. The effects of the external magnetic field are negligible on EC even for the very strong $\bm{B}$. Using the present numerical results obtained, EC is parameterized for $T=(0\sim400)$ MeV with $\sigma=C_\mathrm{EM}(\mathcal{C}_1T+\mathcal{C}_2T^2+\mathcal{C}_3T^3+\cdots)$ and, in this parameterization, the coefficients for $T^2$ and $T^3$ are tiny in comparison to $\mathcal{C}_1$. As a result, I have $\sigma/T\approx (0.46,0.77,1.08,1.39)\,C_\mathrm{EM}$ for $\tau=(0.3,0.5,0.7,0.9)$ fm, respectively. These results are again well compatible with other theoretical estimations. Readers can refer to Ref.~\cite{Nam:2012sg} for more details of the present work. The transport coefficients for quark matter are very important physical quantities for understanding QCD at extreme conditions. In the present work, it was shown that the instanton model reproduced qualitatively well results in comparison to other theoretical results. Other transport coefficients, i.e. the shear and bulk viscosities, are under investigation within the same theoretical framework.   

\section*{Acknowledgments}
The present manuscript was prepared for the proceeding for the international conference \emph{$X$th Quark Confinement and the Hadron Spectrum} (Confinement 10), $8\sim12$ October 2012,  Technische Universit\"at M\"unchen (TUM) Campus Garching, Munich, Germany, and the contents are based on Ref.~\cite{Nam:2012sg}. The author is grateful to Z.~Fodor, N.~Sadooghi, B.~Hiller, and S.~Kim for the fruitful discussions and comments on this work.


\begin{thebibliography}{99}
\bibitem{Voloshin:2008jx} 
  S.~A.~Voloshin [STAR Collaboration],
  Indian J.\ Phys.\  {\bf 85}, 1103 (2011).
\bibitem{Kharzeev:2004ey}
  D.~Kharzeev,
  Phys.\ Lett.\  B {\bf 633}, 260 (2006).
\bibitem{Buividovich:2009wi}
  P.~V.~Buividovich, M.~N.~Chernodub, E.~V.~Luschevskaya and M.~I.~Polikarpov,
  Phys.\ Rev.\  D {\bf 80}, 054503 (2009).
\bibitem{Fukushima:2008xe}
  K.~Fukushima, D.~E.~Kharzeev and H.~J.~Warringa,
  Phys.\ Rev.\  D {\bf 78}, 074033 (2008).
\bibitem{Kharzeev:2010gd} 
  D.~E.~Kharzeev and H.~U.~Yee,
  Phys.\ Rev.\ D {\bf 83}, 085007 (2011).
\bibitem{Nam:2011vn} 
  S.~i.~Nam and C.~W.~Kao,
  Phys.\ Rev.\ D {\bf 83}, 096009 (2011).
\bibitem{Ding:2010ga} 
  H.~T.~Ding, A.~Francis, O.~Kaczmarek, F.~Karsch, E.~Laermann and W.~Soeldner,
  Phys.\ Rev.\ D {\bf 83}, 034504 (2011).
\bibitem{Kerbikov:2012vp} 
  B.~Kerbikov and M.~Andreichikov,
  arXiv:1206.6044 [hep-ph].
\bibitem{BOOKEM}  D.~J.~Evans and G.~P.~Morriss, 
\textit{Statistical Mechanics of Non-Equilibrium Liquids}, 
(Academic Press, London 1990).
\bibitem{Gupta:2003zh} 
  S.~Gupta,
  Phys.\ Lett.\ B {\bf 597}, 57 (2004).
\bibitem{Harada:2006hu} 
  M.~Harada and C.~Sasaki,
  Phys.\ Rev.\ D {\bf 74}, 114006 (2006).
\bibitem{Shuryak:1981ff}
  E.~V.~Shuryak,
  Nucl.\ Phys.\  {\bf B203}, 93 (1982).
\bibitem{Diakonov:1983hh}
  D.~Diakonov, V.~Y.~Petrov,
  Nucl.\ Phys.\  {\bf B245}, 259 (1984).
\bibitem{Diakonov:1988my}
  D.~Diakonov and A.~D.~Mirlin,
  Phys.\ Lett.\  B {\bf 203}, 299 (1988).
\bibitem{Nam:2009nn} 
  S.~i.~Nam,
  J.\ Phys.\ G  {\bf 37}, 075002 (2010).
\bibitem{Schwinger:1951nm}
  J.~S.~Schwinger,
  Phys.\ Rev.\  {\bf 82}, 664 (1951).
\bibitem{Nam:2009jb} 
  S.~i.~Nam,
  Phys.\ Rev.\ D {\bf 80}, 114025 (2009).
\bibitem{Nam:2012sg} 
  S.~i.~Nam,
 Phys.\ Rev.\ D {\bf 86}, 033014 (2012).
\bibitem{Aoki:2006br} 
  Y.~Aoki, Z.~Fodor, S.~D.~Katz and K.~K.~Szabo,
  Phys.\ Lett.\ B {\bf 643}, 46 (2006).
\bibitem{Aoki:2009sc} 
  Y.~Aoki, S.~Borsanyi, S.~Durr, Z.~Fodor, S.~D.~Katz, S.~Krieg and K.~K.~Szabo,
  JHEP {\bf 0906}, 088 (2009).
\bibitem{Borsanyi:2010bp} 
  S.~Borsanyi {\it et al.}  [Wuppertal-Budapest Collaboration],
  JHEP {\bf 1009}, 073 (2010).
\bibitem{Bazavov:2011nk} 
  A.~Bazavov {\it et al.}, 
  Phys.\ Rev.\ D {\bf 85}, 054503 (2012).
\bibitem{Maezawa:2007fd}
  Y.~Maezawa,  S.~Aoki, S.~Ejiri, T.~Hatsuda, N.~Ishii, K.~Kanaya and N.~Ukita,
  J.\ Phys.\ G {\bf 34}, S651 (2007).
\bibitem{Ali Khan:2000iz}
  A.~Ali Khan {\it et al.}  [CP-PACS Collaboration],
  Phys.\ Rev.\  D {\bf 63}, 034502 (2001).
\bibitem{Aarts:2007wj} 
  G.~Aarts, C.~Allton, J.~Foley, S.~Hands and S.~Kim,
  Phys.\ Rev.\ Lett.\  {\bf 99}, 022002 (2007).
\bibitem{Tuchin:2010vs} 
  K.~Tuchin,
  Phys.\ Rev.\ C {\bf 82}, 034904 (2010)
  [Erratum-ibid.\ C {\bf 83}, 039903 (2011)].
\bibitem{Buividovich:2010tn} 
  P.~V.~Buividovich {\it et al.}, 
 {  Phys.\ Rev.\ Lett.\  }{\bf 105}, 132001  (2010).
\bibitem{FernandezFraile:2009mi} 
  D.~Fernandez-Fraile and A.~Gomez Nicola,
  Eur.\ Phys.\ J.\ C {\bf 62}, 37 (2009).

\end{thebibliography}
\end{document}